\documentclass[pdflatex,sn-mathphys-ay]{sn-jnl}

\usepackage{graphicx}
\usepackage{amsmath,amssymb,amsfonts}%
\usepackage{mathrsfs}
\usepackage{xcolor}%
\usepackage{todonotes}
\usepackage{ulem}
\newcommand{\scri}{\mathscr{I}}

\begin{document}

\title[Hyperboloidal Foliations]{Topical Collection— Hyperboloidal Foliations in the Era of Gravitational-Wave Astronomy: From Mathematical Relativity to Astrophysics}

\author*[1]{\fnm{David} \sur{Hilditch}}\email{david.hilditch@tecnico.ulisboa.pt}
\author*[2]{\fnm{Rodrigo} \sur{Panosso Macedo}}\email{rodrigo.macedo@nbi.ku.dk}
\author*[1,3]{\fnm{Alex} \sur{Vañó-Viñuales}}\email{alex.vano@uib.es}
\author*[4]{\fnm{Anıl} \sur{Zenginoğlu}}\email{anil@umd.edu}

\affil[1]{\orgdiv{CENTRA, Departamento de Física}, \orgname{Instituto Superior Técnico IST, Universidade de Lisboa UL}, \orgaddress{\street{Avenida Rovisco Pais 1}, \city{Lisboa}, \postcode{1049}, \country{Portugal}}}

\affil[2]{\orgdiv{Centre of Gravity}, \orgname{Niels Bohr Institute}, \orgaddress{\street{Blegdamsvej 17}, \city{Copenhagen}, \postcode{2100}, \country{Denmark}}}

\affil[3]{\orgdiv{Departament de Física}, \orgname{Universitat de les Illes Balears, IAC3}, \orgaddress{\street{Carretera Valldemossa km 7.5}, \city{Palma}, \postcode{E-07122}, \country{Spain}}}

\affil[4]{\orgdiv{Institute for Physical Science \& Technology}, \orgname{University of Maryland}, \orgaddress{\city{College Park}, \postcode{MD 20742-2431}, \country{USA}}}

\maketitle

Gravitational-wave astronomy has blossomed in the past few years, giving us a new channel to observe the Universe and to empirically study strong-field gravity. The detectors of the LIGO-Virgo-KAGRA network now routinely probe regions where spacetime is strongly curved and dynamical. Theoretical models describe the emission and propagation of gravitational radiation from distant astrophysical sources to these detectors. The natural geometric construction to formulate such radiation is future null infinity, $\scri^{+}$: the idealized boundary reached by outgoing null rays. A faithful computation of gravitational waves therefore requires a numerical framework that includes $\scri^{+}$ rather than hiding it behind artificial outer boundaries. One can reach null infinity by timelike, spacelike, or null hypersurfaces. From the point of view of an initial value problem, only null and spacelike hypersurfaces are suitable. 

In relativity, an idealized observer propagates along future null infinity and measures gravitational radiation emitted by distant sources. If we consider time hypersurfaces constructed by such far away observers, we are led to the notion of hyperboloidal foliations: spacelike hypersurfaces representing a moment of time for idealized observers. 
By slicing spacetime with spacelike hypersurfaces that asymptote to $\scri^{+}$, hyperboloidal foliations satisfy two requirements: they retain the well-posed and flexible initial-value structure familiar from Cauchy formulations, and they provide access to the radiative zone.

Over the past two decades, hyperboloidal ideas have migrated from mathematical relativity into numerical relativity and astrophysical modeling. Today they are used in a variety of contexts, from linear perturbations on fixed backgrounds to fully nonlinear Einstein equations. The hyperboloidal approach has been applied to study gravitational waves from binary black hole mergers. 

To bring these diverse developments together, we organized the workshop “Infinity on a Gridshell” in Copenhagen in 2023. The meeting gathered experts in conformal geometry, partial-differential equations, numerical analysis, and gravitational-wave astronomy.  Talks and discussion sessions ranged from global existence theorems, through discontinuous Galerkin schemes for self-force computations, to practical recipes for attaching hyperboloidal layers to existing binary-merger codes.  The exchange between mathematical and numerical relativists made it clear that progress on the hyperboloidal initial-value problem requires cross-disciplinary collaboration.

The topical collection, “Hyperboloidal Foliations in the Era of Gravitational-Wave Astronomy: From Mathematical Relativity to Astrophysics,” grew out of those interactions. Contributors were invited to expand their Copenhagen presentations into full articles, or submit new research motivated by the discussions. The result is a snapshot of a field in rapid development: papers that solidify the linear theory on fixed backgrounds, push numerical techniques to spectral accuracy at null infinity, clarify the role of conserved charges and regularity at $\scri^{+}$, and take further steps toward fully nonlinear hyperboloidal evolutions without symmetry.

We hope that the collection will serve both as a reference for researchers already working with hyperboloidal methods and as an accessible entry point for newcomers motivated by recent developments. In this editorial, we group the contributions thematically and highlight how each one fits into the landscape from well-understood linear cases to the challenging nonlinear regime.

\section{Black Hole Perturbations}

Hyperboloidal methods are best developed for perturbations on background spacetimes. They serve as a tool for investigating the properties of wave propagation in black-hole spacetimes. Several papers in this collection exploit this setting to deepen our understanding of wave dynamics and radiation extraction at null infinity.


\cite{BessonJaramillo2025} study QNM expansions in black hole perturbation theory using a spectral approach on hyperboloidal slices. They introduce a hyperboloidal Keldysh scheme to formulate the QNM analysis as a non-selfadjoint spectral problem. The radiating boundary conditions at infinity are automatically built in to geometry, and the QNMs can be treated via the spectral theory of the operator's resolvent. Using Keldysh's theorem on bi-orthogonal systems, Besson and Jaramillo construct an asymptotic expansion of the propagator in terms of resonant modes, yielding a spectral version of the classical Lax-Phillips scattering expansion. 
They clarify that while a Hilbert space inner product is not needed to expand the solution in resonant modes at null infinity, the choice of an inner product becomes critical when defining the excitation coefficients in the bulk. The hyperboloidal Keldysh expansion is highly accurate and efficient, able to reproduce known features of black hole dynamics even beyond its formal validity. Surprisingly, the method even captures the late-time power-law tail in Schwarzschild spacetime. 
The authors push their analysis to compute second-order QNMs and derive a Weyl law for counting QNM frequencies in various asymptotic geometries. These developments have clear relevance for gravitational-wave astronomy: QNMs are the spectral fingerprint of black hole spacetimes, and understanding their completeness and mode-sum convergence is vital for interpreting ringdown signals.

\cite{MinucciMacedo2025} make a beautiful connection between spacetime geometry and special functions. Their work demonstrates that the confluent Heun functions providing the formal solutions of the radial Teukolsky equation have a geometric meaning: each canonical form of the confluent Heun equation corresponds to a different choice of slicing (what they call ``Heun slices''). The local behaviour of these functions near their regular and irregular singular points is a direct reflection of the black hole's global structure, linking the solutions to different causal regions of the Kerr spacetime. Their analysis clarifies how homotopic transformations of the Heun functions amount to re-slicing the manifold and thus provides a geometric framework for understanding wave propagation and scattering from the horizon to infinity.

\cite{Leather2025}  computes the gravitational self-force in the Lorenz gauge using hyperboloidal slicing and multi-domain spectral methods. 
This work uses the minimal gauge for the hyperboloidal slicing and the Lorenz gauge for gravitational perturbations. The computation includes radiative fluxes, the Detweiler redshift, and self-force corrections for a quasicircular orbit. Leather's approach provides high-accuracy gravitational self-force, resulting in substantial computational savings. The application of hyperboloidal foliations is essential for extending self-force computations to second order.


\cite{VishalFieldEtAl2025} present a discontinuous Galerkin (DG) method that achieves superconvergence on hyperboloidal slices. They solve the scalar Teukolsky equation, modeling a perturbation on a Kerr background, with a point-particle source. Standard high-order DG methods exhibit spectral convergence, with errors decreasing with the power of the polynomial degree of expansion. Remarkably, the authors identify conditions under which the DG scheme becomes superconvergent, 
doubling the formal order of accuracy for phase errors and dispersion. The authors demonstrate this superconvergence property for their hyperboloidal DG scheme and show that it persists even when employing hyperboloidal layer compactification. This numerical method allows for long and accurate evolutions as numerical dispersion and dissipation are drastically reduced and the computation of the gravitational dissipative self-force has high accuracy. Such techniques are essential for modeling extreme mass-ratio inspirals, where the small mass ratio requires tracking many orbital cycles with minimal phase errors. The superconvergence property of the hyperboloidal DG scheme is a significant advantage in this context.
 
\cite{BishoyiSabharwalKhanna2025} investigate the existence of non-axisymmetric gravitational hair on extremal Kerr black holes. They write the Teukolsky equation in horizon-penetrating hyperboloidal coordinates thereby eliminating artificial outer boundaries and enabling direct waveform extraction at infinity. Building on Aretakis's conserved charges for axisymmetric perturbations, the authors introduce a conjectured non-axisymmetric horizon charge defined via the transverse derivative of the invariant Beetle-Burko scalar $\xi = \psi_{0}\,\psi_{4}$. Through long-time, high-resolution evolutions of spin-weighted quadrupolar and octupolar modes, they demonstrate that $\xi$ decays as an inverse power of advanced time along the horizon, but its first transverse derivative approaches a nonzero constant, numerically indicating the presence of gravitational hair. They further adapt Ori's late-time expansion to show that this horizon charge can be extracted from the radial fall-off of $\xi$ at finite radii, thus establishing a concrete link between near-horizon structure and asymptotic observables. This study not only provides numerical evidence for non-axisymmetric hair in extremal Kerr spacetimes but also demonstrates the effectiveness of hyperboloidal compactification for resolving both near-horizon dynamics and far-field radiation with high accuracy.

\cite{RaczToth2024} carry out a comprehensive numerical investigation of late-time tails of wave solutions in Kerr spacetime. They focus on the Fackerell-Ipser equation, a wave equation governing a spin-0 component of the electromagnetic field in Kerr spacetime. Using horizon-penetrating, hyperboloidal coordinates, the authors evolve spatially compact, pure multipole perturbations, and monitor the solution both at the event horizon and future null infinity. The authors find that for a wide range of initial configurations, the field at late times approaches either a constant static solution or zero, after an exponential QNM decay and a polynomial fall-off. This work extends the classic Price's law, originally derived for scalar perturbations on a Schwarzschild spacetime, to a more general context. 

The contributions on linear perturbations and fixed backgrounds show that hyperboloidal methods have reached a level of maturity where both analytical and numerical approaches agree and complement each other. We now have exact solutions and spectral expansions validating the hyperboloidal approach, as well as highly accurate simulations that compute waveforms all the way to null infinity.

\section{Geometry, Asymptotic Structure and Initial Data}

This section gathers contributions that lay the geometrical groundwork for the hyperboloidal approach. The contributions address how to construct suitable hyperboloidal coordinates, regular hyperboloidal initial data, and invariant quantities at infinity. One recurring question in mathematical relativity is the smoothness of the compactified spacetime at null infinity. We would like the conformal metric $g_{ab} = \Omega^2 \tilde g_{ab}$ (where $\tilde g_{ab}$ is the physical metric and $\Omega$ vanishes at $\scri$) to be regular at the conformal boundary $\{\Omega=0\}$. Smoothness ensures that physical quantities like the Bondi energy and radiation flux are well-defined and finite at infinity. The first two contributions below focus on this asymptotic smoothness and the associated conserved quantities.

\cite{CsukasRacz2025} examine hyperboloidal data for Einstein's equations. Such data tend to have polyhomogeneous (log-containing) expansions near infinity, which implies the spacetime is not $C^\infty$ at $\scri$ but only formally smooth to a certain order.  Building on earlier results of Andersson and Chruściel, the authors show that requiring a well-defined Bondi mass selects a subclass of solutions to the conformal constraint equations that guarantee a smooth conformal boundary. They show that if a hyperboloidal initial data set has a well-defined Bondi mass, then that initial data evolves to a spacetime with a smooth conformal boundary. They then strengthen this condition: if both Bondi mass and angular momentum are well-defined (and some mild fall-off conditions hold), then the solution of the parabolic-hyperbolic form of the constraint equations is free of logarithmic terms. The paper includes numerical examples of such initial data in the vicinity of a Kerr black hole. This study implies that the logarithmic divergences are not inevitable. Many analytic formulas in gravitational wave science, such as the Bondi mass loss, or the peeling properties of curvature, assume a certain degree of smoothness at infinity. The authors give a clear foundation for when those assumptions hold, and provide practical guidance for setting up hyperboloidal initial data in numerical codes that do not include logarithmic terms. 

\cite{SancassaniVelu2025} study how physical conserved quantities, such as energy and linear momentum, evolve along hyperboloidal slices. 
The authors introduce the notion of E-P chargeability to describe initial data that admit well-defined energy ($E$) and linear momentum ($P$). They prove that if the initial data is E-P chargeable, then under Einstein's evolution this property is preserved with an appropriate choice of time coordinate. Using a hyperboloidal foliation, the authors derive the flux formulas for energy and linear momentum directly from the Einstein evolution equations, recovering the standard Bondi-Sachs energy-loss and momentum-loss formulas, but now derived under weaker asymptotic assumptions than usually required. In particular, their approach does not assume a full conformal compactification of spacetime; instead, it operates at the level of the initial 3-geometry and its embedding in spacetime. 

\cite{RossettiVanoVinuales2025} adapt the hyperboloidal method to cosmological spacetimes. The spacetimes they consider are expanding Friedmann--Lemaître--Robertson--Walker (FLRW) spacetimes with a time-dependent scale factor. By introducing both conformal time and time-dependent height functions, they construct compactified hyperboloidal slices that naturally intersect $\scri$ for different signs of spatial curvature. Their numerical experiments confirm the decay rates of linear waves predicted by analytical estimates in a wide class of FLRW models. Crucially for gravitational-wave astronomy, this work demonstrates that hyperboloidal foliations can be adapted to evolve cosmological backgrounds. Going beyond linear studies, they present numerical evidence for small-data global existence of semi-linear wave equations under a generalized null condition in decelerating universes, and they identify a threshold in the expansion rate below which solutions blow up in finite time. 

\cite{Zenginoglu2025} provides a unifying framework for horizon-penetrating, hyperboloidal coordinates. This work points out the deep connection between the familiar horizon-penetrating coordinates (regular across black-hole event horizons and cosmological horizons) and hyperboloidal coordinates (which approach null infinity) as special cases of regular null-transverse foliations. Beginning with a review of classical horizon-penetrating slices (Painlevé--Gullstrand, Eddington--Finkelstein) and global extensions (Kruskal--Szekeres, Penrose diagrams), the author discusses a general height-function ansatz in stationary, spherically symmetric spacetimes that smoothly crosses both the event horizon and future null infinity within a single time-translation--invariant slicing. As these coordinates bridge between null hypersurfaces, the term bridging foliation is introduced. Several examples are given to illustrate this framework, such as source-adapted hyperboloidal slicings and generalized Fefferman--Graham--Bondi coordinates. By demonstrating that both horizon-penetrating and hyperboloidal slicings are just regular choices of time that extend through null horizons, this study provides a theoretical framework for regular choices of coordinates and practical guidelines for numerical simulations. 

\cite{ValienteKroonDaSilva2024} analyze the 1+1-dimensional d'Alembert solution in a hyperboloidal slicing. By deriving the exact solution of the flat-space wave equation in hyperboloidal coordinates, they provide intuition for wave propagation on hyperboloidal time slices. Notably, their hyperboloidal d'Alembert solution explains an apparently anomalous effect seen in previous numerical studies: a permanent displacement of the field after a wave pulse has passed. In the standard setting, an initial perturbation with compact support would leave no memory after it disperses. However, in hyperboloidal coordinates, numerical experiments had observed a non-zero offset. The authors demonstrate that this permanent displacement is not a numerical artifact but rather an intrinsic feature of wave propagation in hyperboloidal foliations, arising from how outgoing radiation is represented on slices that reach null infinity. 

These contributions considerably further our understanding of the regularity of null infinity, the desired properties of the hyperboloidal foliations in use, and spacetimes where the hyperboloidal infrastructure can be applied. This knowledge is not only highly valuable {\it per se}, but it is also of utmost importance for the treatment of null infinity in the nonlinear case.

\section{Advances Toward the Full Nonlinear Problem}

Extending the successes of the hyperboloidal approach to the full, nonlinear Einstein equations has remained an open problem. The ultimate goal is to perform hyperboloidal evolutions of the fully nonlinear Einstein equations describing astrophysical scenarios, such as binary black hole mergers. In the fully nonlinear regime, one encounters all the difficulties of regular numerical relativity (gauge issues, stability, high computational cost) in addition to new ones introduced by the conformal compactification (regularization of the equations at $\scri$). While this problem remains unresolved, the contributions in this collection demonstrate significant progress.

\cite{PetersonGasperinVV2025} perform numerical evolutions of the linearized conformal Einstein field equations on a flat background using the inversion symmetry of Minkowski spacetime. This work investigates the scri-fixing technique for the conformal Einstein equations. In scri-fixing, one chooses gauge conditions such that future null infinity lies at a fixed coordinate location that can then be mapped to the outer boundary of the numerical grid. The authors evolve the linearized conformal Einstein equations 
and verify that the numerical solution remains regular up to the boundary at null infinity. They compare runs with the scri-fixing gauge versus a more traditional gauge choice. The results confirm that scri-fixing works as intended at the linear level: the formally singular terms that often plague hyperboloidal evolution in standard coordinates are regularized, and one can stably propagate waves to $\scri$ without loss of accuracy. This study shows that conformal field equations (introduced by Friedrich for the mathematical analysis of the Einstein equations) can be leveraged for numerical simulations with scri-fixing.

\cite{CamdenFrauendiener2025} implement Friedrich's generalised conformal field equations in the conformal Gauss gauge. In this setup, a bounded initial hypersurface with an expanding timelike boundary naturally intersects future null infinity, thereby becoming hyperboloidal during time evolution. Their public COFFEE code combines pseudo-spectral angular discretisation using spin-weighted spherical harmonics with fourth-order radial finite differences, enabling direct computation of Bondi--Sachs mass loss, the onset and frequencies of quasi-normal ringing, and the preservation of Newman--Penrose charges. This work provides a numerical link between finite and asymptotic observables and validates key formulae to high accuracy.

\cite{Duarte2025} derives rigorous, uniform energy bounds for a prototypical system that mimics the nonlinear structure of the Einstein equations in generalized harmonic gauge.  In the first part of the work, Duarte shows that when the ``ugly" field is rescaled by the radial coordinate and its inhomogeneities vanish, its evolution equation reduces to the familiar ``good" wave equation, allowing one to import standard $L^2$ conservation laws and the Klainerman--Sobolev decay estimates on Cauchy slices.  He then uses this connection to prove that the incoming-null-derivative of the ugly field decays one order faster than in the good case. In the second part, Duarte tackles the general case of nonzero source terms and arbitrary decay parameter $p$.  By performing a first-order reduction, and compactifying the radial coordinate, he casts the system into a symmetric hyperbolic form on hyperboloidal slices. A Gronwall-type argument then establishes that the energy of the compactified variables remains uniformly bounded under evolution. The paper provides an approach for extending energy estimates to the full first-order, compactified Einstein equations in generalized harmonic gauge. 

\cite{VanoVinualesValente2024} systematically explore the gauge conditions via 4D reference metrics for hyperboloidal slicing in spherical symmetry. One major hurdle in hyperboloidal evolution is choosing coordinate gauge functions (such as lapse and shift or their generalizations) that keep the computation stable and regular. The authors propose to build these gauge source functions from an analytically chosen reference metric that encodes the desired asymptotic behavior. Using the height-function method, they generate a family of reference metrics for Minkowski spacetime that yield suitable hyperboloidal slices. They test ten different choices of height functions (including three with a hyperboloidal layer) and solve the full Einstein equations in spherical symmetry for each. They demonstrate, for the first time, stable long-term integrations of the Einstein equations with hyperboloidal layers. In  hyperboloidal layer gauge, the interior part of the slice behaves like a standard Cauchy slice, and only after a certain radius does the slice approach future null infinity. This gauge is attractive for hybrid evolution schemes combining well-tested Cauchy codes with hyperboloidal codes. Although performed in spherical symmetry, the lessons from this study are potentially instructive for future 3D codes. If these techniques generalize to 3D, one could perform simulations of merging binaries in puncture coordinates and directly output waveforms at infinity with high accuracy, improving the modeling of signals for gravitational-wave detectors.

\cite{LeFlochMa2024} present a nonlinear hyperboloidal global existence result in modified gravity. They employ the Euclidean-hyperboloidal foliation method, previously developed for handling the Einstein equations with scalar fields, and apply it to $f(R)$ gravity. By casting the modified field equations into a coupled system of nonlinear wave and Klein-Gordon equations, LeFloch and Ma prove a nonlinear stability theorem for Minkowski spacetime in this setting. The authors formulate the problem in conformal wave gauge and use energy estimates tailored to hyperboloidal slices. They show that an initial data set for $f(R)$ gravity sufficiently close to Minkowski data evolves to a global solution that approaches Minkowski spacetime at infinity towards both spatial and null infinity. This is a classic global existence and stability result, analogous to the famous Minkowski stability result by Christodoulou and Klainerman. This paper demonstrates an alternative approach to global existence in hyperboloidal foliations applied to higher-order gravity theories. 

\section*{Acknowledgements}
Thanks are due to to Frank Schulz for the support during the creation of the collection, to Vitor Cardoso for his support for in setting up the workshop, and of course to all of the participants.

We gratefully acknowledge support for the workshop from the VILLUM Foundation (grant no. VIL37766) and the DNRF Chair program (grant no. DNRF162) by the Danish National Research Foundation.

RPM further acknowledge support from he European Union's Horizon 2020 research and innovation programme under the Marie Sklodowska-Curie grant agreement No 101007855 and No 101131233. The Center of Gravity is a Center of Excellence funded by the Danish National Research Foundation under grant No. 184.

DMH acknowledges support from FCT (Portugal) projects UIDB/00099/2020 and UIDP/00099/2020 and PeX-FCT (Portugal) program 2023.12549.PEX.

AVV also thanks FCT for support from project UIDB/00099/2020, as well as funding with DOI 10.54499/DL57/2016/CP1384/CT0090. Graciously acknowledged is also the support of the Spanish grants PID2022-138626NB-I00, RED2022-134204-E, RED2022-134411-T, funded by MICIU/AEI/10.13039/501100011033; and the European-supported Balearic Islands regional projects SINCO2022/6719 and SINCO2022/18146.

AZ acknowledges support by the National Science Foundation under Grant No. 2309084.

\section*{Declarations}

\noindent\textbf{Competing interests:} The authors declare no competing interests.
\vspace{3mm}

\noindent\textbf{Publisher's Note:} Springer Nature remains neutral with regard to jurisdictional claims in published maps and institutional affiliations.

\bigskip

\bibliography{refs}

\end{document}